\documentclass[aps,prd,twocolumn,groupedaddress,showpacs,showkeys,preprintnumbers]{revtex4}

\usepackage{amsmath,amssymb,graphicx}

\newcommand{\fsize}{0.45\textwidth}
\newcommand{\EW}{E\"ot-Wash}
\newcommand{\ew}{EW}
\newcommand{\B}[1]{\textbf{#1}}
\newcommand{\mc}[1]{\mathcal{#1}}
\newcommand{\mat}[1]{\overline{\textbf{#1}}}
\newcommand{\halfint}{\int_{0}^{\infty}}
\newcommand{\fullint}{\int_{-\infty}^{\infty}}

\begin{document}


\title{Optimal Determination of the Equilibrium Displacement of a Damped Harmonic Oscillator in the Presence of Thermal Noise}


\author{Michael W. Moore}
\author{Jason H. Steffen}
\email[]{jsteffen@astro.washington.edu}
\author{Paul E. Boynton}
\affiliation{University of Washington, Department of Physics}


\date{\today}

\begin{abstract}
Using a matched filter technique, we derive the minimum variance, unbiased estimator for the equilibrium displacement of a damped harmonic oscillator in thermal equilibrium when interactions with the thermal bath are the leading source of noise.  We compare the variance in this optimal estimator with the variance in other, commonly used estimators in the presence of pure thermal noise and pure white noise.  We also compare the variance in these estimators for a mixture of white and thermal noise.  This result has implications for experimental design and the collection and analysis of data.
\end{abstract}

\pacs{02.50, 02.60}
\keywords{Thermal Noise, Damped Harmonic Oscillator, Power Spectrum, Torsion Device, Data Analysis, Random Walk}

\maketitle

\section{Introduction \label{penalty}}

The torsion pendulum is currently used in a number of experimental programs to test theories of gravity (\cite{fis} and references therein).  This work involves the detection of extremely small torques and requires the experimentalist to design measurements whose precision approaches the fundamental limit posed by thermal noise prescribed by optimal (minimum variance, unbiased) statistical estimation techniques.  This requirement is familiar to the community engaged in these studies, and those readers will immediately ask why we need still another treatment of thermal noise on a damped harmonic oscillator.  To answer this question, we begin with a simple example that illustrates a major shortcoming in customary methodologies.

Consider a linear oscillator in thermal equilibrium with a heat bath at absolute temperature, $T$.  The equilibrium displacement of the pendulum, $c$, can be estimated by measuring the instantaneous displacement of the oscillator, $x(t)$, at $t=0$
\begin{equation}
\hat{c}_{ins}=x(0)
\end{equation}
where the circumflex indicates a parameter estimate.  The ensemble of such estimates is a random variable (the estimator), and it is represented by $\hat{C}_{ins}$.  We will use the convention that a capital letter represents an ensemble and a lower-case letter represents a realization of the ensemble.  The equipartition theorem prescribes the variance of the instantaneous estimator 
\begin{equation}
\text{var}( \hat{C}_{ins} ) = \frac{k_bT}{\kappa} \equiv \sigma^2 ,
\label{cinst}
\end{equation}
where $k_b$ is the Boltzmann constant and $\kappa$ is the torsional spring constant.

If the data used in a parameter estimate is a continuous time series, $x(t)$, one can define another estimator that has smaller variance than the instantaneous estimator.  A familiar approach, the ``boxcar'' estimate, is an average of the displacement of the oscillator over the time series starting at $t=0$ and ending at $t=\tau$,
\begin{equation}
\hat{c}_{box}=\frac{1}{\tau}\int_0^{\tau}x(t)dt.
\label{cboxcar}
\end{equation}
For the case of an oscillator dominated by thermal noise, one can calculate the variance of the boxcar estimator using the Fourier methods presented later in this article
\begin{equation}
\begin{split}
&\text{var}(\hat{C}_{box}) = \frac{2 \sigma^2}{\omega \omega_0^4 \tau^2} \Bigl( -3\gamma^2\omega + \omega^3 + 2\gamma^3\omega\tau + 2 \gamma \omega^3 \tau \\
&+ e^{-\gamma\tau} \bigl( \left( 3\gamma^2\omega - \omega^3 \right) \cos(\omega\tau) + \left( \gamma^3 - 3\gamma\omega^2 \right) \sin(\omega\tau) \bigr) \Bigr)
\label{varboxcar}
\end{split}
\end{equation}
where $\omega_0=\sqrt{\kappa/m}$ is the undamped oscillation frequency, $\gamma$ is the decay coefficient, and $\omega=\sqrt{\omega_0^2-\gamma^2}$ is the damped oscillation frequency.  The quality factor, $Q=\omega_r/(2\gamma)$, is traditionally defined in terms of the resonant frequency, $\omega_r=\sqrt{\omega_0^2-2\gamma^2}$, but to simplify equations appearing later, it is convenient to define an alternate quality factor, $Q_0=\omega_0/(2\gamma)$, in terms of the undamped frequency.  Similarly, we state the duration of the data sample in units of undamped oscillation periods, $N=\omega_0 \tau/(2\pi)$.

\begin{figure}
\begin{center}
\includegraphics[width=\fsize]{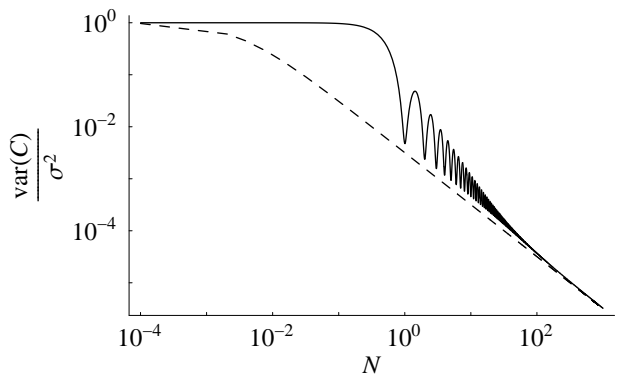}
\caption{The solid curve is the variance of the boxcar estimator of the equilibrium displacement, in units of the variance of the instantaneous estimator, versus the duration of the data sample in undamped oscillator periods.  The oscillator has a quality factor of $Q_0=50$.  The dashed curve is a plot of the variance of the optimal estimator, also in terms of the variance of the instantaneous estimator and for $Q_0=50$. \label{figure01}}
\end{center}
\end{figure}

The solid curve in Figure \ref{figure01} shows the logarithm of $\text{var}( \hat{C}_{box})/\sigma^2$ versus the logarithm of $N$ for an oscillator of quality factor, $Q_0=50$.  It is apparent that the boxcar estimate is not optimal for the thermal-noise-dominated pendulum because the variance does not decrease monotonically with increasing sample duration $\tau$.  Adding more data cannot degrade the optimal estimate of a parameter.  What then \textit{is} the optimal estimate, $\hat{c}_{op}$?

We assert that the dashed curve in Figure \ref{figure01} represents the variance of the optimal estimator, $\text{var}( \hat{C}_{op})/\sigma^2$.  This curve is qualitatively consistent with what one would expect from an optimal estimator---it does decrease monotonically with increasing $\tau$, and it lies on or below the boxcar estimator for all values of $\tau$.  In this paper, we derive a closed-form expression for $\hat{c}_{op}$ under fairly general assumptions.  The point of presenting this example is that up to now a solution to this basic estimation problem has not appeared in the experimental literature.  This is the gap we wish to fill with this article.

It is somewhat surprising that as we approach the $100^{th}$ anniversary of Einstein's seminal work on Brownian motion~\cite{ein} there remain several, arguably canonical, questions whose answers are not widely known in the physics community.  Aside from the large number of people who have looked at the problem, several notable minds have studied Brownian motion on a damped harmonic oscillator.  For example, in Chandrasekhar's 1943 \textit{Reviews of Modern Physics} article, ``Stochastic Problems in Physics and Astronomy''~\cite{cha}, one learns that given the initial displacement and velocity, $x_0$ and $v_0$, of a damped harmonic oscillator at $t=0$, the probability distribution function for the displacement $x$ at time $t>0$ is
\begin{widetext}
\begin{equation}
\begin{split}
&W(x,t;x_0,v_0)=\left(\frac{m}{4\pi\beta k_b T\int_0^t\psi^2(\psi)d\psi}\right)^{\frac{1}{2}}\times \\
&\exp\left( -\frac{m\left(x-x_0e^{-\beta t/2}\left(\cosh\left(\frac{\beta_1 t}{2}\right)+\frac{\beta}{\beta_1}\sinh \left(\frac{\beta_1 t}{2}\right)\right)-\frac{2v_0}{\beta_1}e^{-\beta t/2}\sinh\left(\frac{\beta_1 t}{2}\right)\right)^2}{4\beta k_b T \int_0^t \psi^2(\psi)d\psi}\right)
\end{split}
\label{chandra}
\end{equation}
where
\begin{equation}
\begin{split}
&\int_0^t \psi^2(\psi)d\psi = \frac{1}{2\omega_0^2\beta} - \frac{e^{-\beta t}}{2\omega_0^2\beta_1^2\beta}\left( 2\beta^2 \sinh^2\left(\frac{\beta_1 t}{2}\right)+\beta \beta_1 \sinh(\beta_1 t) + \beta_1^2 \right)\\
&\beta_1=\left(\beta^2 - 4\omega_0^2\right)^{\frac{1}{2}}\\
&\beta=2\gamma.
\end{split}
\end{equation}
\end{widetext}
This result completely describes the stochastic time evolution of a damped harmonic oscillator in contact with a heat bath.  For a theorist, the problem is solved.  And in a limited sense, this assessment is correct.  Probability theory, which produces equation (\ref{chandra}), attempts to characterize the measured values one would obtain given the parameters of the system.  A statistical approach, on the other hand, is concerned with the inverse problem: to determine a measurement/inference scheme that provides optimal estimates for the relevant parameters.  As such, a statistical characterization is of keen interest for experimentalists because it provides insight into both the design of the experiment and analysis of the experimental data.

To answer typical statistical questions that experimentalists wish to ask regarding the damped harmonic oscillator, the autocovariance function of the stationary thermal noise ensemble provides sufficient information.  Its form
\begin{equation}
\begin{split}
\bigl< \delta & X_{th}(t) \delta X_{th}(t+\Delta t) \bigr>\\
&=\sigma^2e^{-\gamma |\Delta t|}\left(\cos(\omega|\Delta t|)+\left(\frac{\gamma}{\omega} \right)\sin(\omega|\Delta t|) \right)
\label{autocovar}
\end{split}
\end{equation}
is much simpler than (\ref{chandra}) because of the time-translation invariance of the stationary noise ensemble.  Using either (\ref{chandra}) or (\ref{autocovar}) one can calculate the variance of a particular estimator, but neither equation alone yields the minimum variance, unbiased estimator.

The question of optimal estimation has been studied extensively.  In particular, because of applications to radar, optimal filter theory was intensely developed during World War II.  For stationary noise processes, analyzing statistical estimation in the Fourier basis greatly simplifies the problem because the noise fluctuations in various Fourier components are not correlated one with another.  The power spectrum of the thermal noise ensemble corresponding to the autocovariace shown in (\ref{autocovar}),
\begin{equation}
S\left[ \delta X_{th} \right] = \frac{8\sigma^2 \gamma \omega_0^2}{\left( \left(2 \pi \nu \right)^2 - \omega_0^2\right)^2 + \left( 4 \pi \gamma \nu \right)^2},
\label{specpow}
\end{equation}
is a Fourier representation containing the same information.  Basically, optimal filter theory states that the optimal estimate is a weighted least-squares sum in Fourier space with the weights being determined by the signal-to-noise ratios of the various Fourier components.

No detailed mathematical derivation is needed to obtain an optimal estimator for the constant $c$.  Since the deflection parameterized by $c$ has only a zero-frequency Fourier component, the optimal estimator, $\hat{C}_{op}$, will also have only a zero-frequency component. Thus, for any stationary noise process, optimal filter theory dictates that $\hat{C}_{op}$ is the boxcar estimator.  Yet, according to the discussion of Figure \ref{figure01}, it would appear that optimal filter theory produces the wrong result.

This discrepancy arises because there are certain assumptions that must be satisfied in order for optimal filter theory to be valid.  Chief among these assumptions is that the discrete Fourier components of the noise from finite duration data samples should be a good approximation to the continuous Fourier components of the noise from infinite duration data samples.  For $N \gg Q_0$, where $N$ is the number of undamped oscillation periods, this requirement is satisfied, and Figure \ref{figure01} shows that $\text{var}(\hat{C}_{box})$ does indeed approach $\text{var}(\hat{C}_{op})$ for $N \gg 50$.  For $N < Q_0$, the narrowband thermal noise of a damped harmonic oscillator does not satisfy the above requirement, and optimal filter theory is not valid in that regime.  When performing atomic force cantilever experiments for which the resonant frequency is measured in kiloHertz and the characteristic damping time, $1/\gamma$, is measured in seconds, waiting for $N \gg Q_0$ is a realistic possibility.   For torsion balance experiments with milliHertz resonant frequencies and characteristic damping times of weeks, however, waiting for $N \gg Q_0$ in order to simplify the data analysis is clearly impractical.  The majority of torsion balance experiments are conducted in the ``bumpy'' regime of the boxcar estimator in Figure \ref{figure01} where optimal filter theory fails most miserably.

Insight to the character of this problem has been suggested by Priestly~\cite{pri} in the treatment of a related question\footnote{In \textit{Spectral Analysis and Time Series}, Priestley shows from the work of Weiner and Kolmogorov that, given the displacement of the oscillator from $t=-\infty$ to $t=0$, the best estimate of the displacement of the oscillator at any time in the future is to extrapolate the damped oscillation forward to that time.  This result is the answer to a related, but fundamentally different question.  It predicts the best estimate of the displacement given the history of the oscillator.  It does not give the best estimate of the parameter corresponding to the equilibrium displacement.}.   Moreover, a formal solution was posed by Grenander~\cite{gre}, but we find that it provides the physicist with neither a great deal of physical insight nor a straightforward means of translating the results into equations involving measurements and the physical model.  We therefore construct our derivation with methods and tools more familiar to the experimental community and refer the interested reader to Grenander for a rigorous mathematical development.

Having set out the question, we now present the answer.  The optimal estimate of deflection is
\begin{equation}
\hat{c}_{op}=\frac{x_i+x_f+Q_0\omega_0\tau x_m + Q_0\left(\frac{v_f-v_i}{\omega_0}\right)}{2+Q_0\omega_0\tau},
\label{copth}
\end{equation}
where $x_i=x(0)$, $x_f=x(\tau)$, $v_i=v(0)$, $v_f=v(\tau)$, and $x_m$ is the boxcar estimate (\ref{cboxcar}).  The variance of the corresponding estimator
\begin{equation}
\text{var}(\hat{C}_{op})=\frac{2\sigma^2}{2+Q_0\omega_0 \tau}
\label{varcopth}
\end{equation}
is the smallest possible for an unbiased estimator of $c$ constrained to using data of duration $\tau$.  The dashed line in Figure \ref{figure01} is a plot of (\ref{varcopth}) for $Q_0=50$.  The derivation and discussion of (\ref{copth}) and (\ref{varcopth}) are the major topics of this paper.  It is not immediately apparent how these results follow from either (\ref{autocovar}) or (\ref{specpow}), but the simplicity of (\ref{varcopth}) implies that symmetries and appropriate transformations streamline the solution.  The complexity of (\ref{chandra}) suggests the problem can become difficult if these symmetries are ignored.

In section \ref{linear} we present the estimation of linear parameters in the presence of noise and show how to calculate the variance in the parameter estimators using the spectral power density of the noise process in the fundamental observable.  We then derive the minimum variance, unbiased estimator of the equilibrium displacement of the torsion pendulum in the presence of white noise and thermal noise in section \ref{derivation}.  Section \ref{physical} examines the effects of multiple noise processes on the variance of different estimators.  In particular we examine a superposition of white noise and thermal noise as well as transients caused by nonthermal disturbances to the oscillator.

\section{\label{linear}Linear Parameters}

\subsection{Estimating Linear Parameters}

A realization of data $x(t)$ is a combination of the physical signal, denoted $x(t;\rho)$ as explained shortly, and a term representing additive noise
\begin{equation}
x(t) = x(t;\rho) + \delta x(t).
\label{datadef}
\end{equation}
We use the convention that a Greek letter $\rho$ represents the physical value of the corresponding parameter $p$ upon which the signal depends.  A linear parameter is estimated by projecting a realization of the data $x(t)$ onto an estimating function $e_{\hat{p}}(t)$
\begin{equation} \label{paramestimate}
\hat{p} = \fullint e_{\hat{p}}(t) x(t) dt.
\end{equation}
In the absence of noise, the data matches the physical signal and an unbiased estimating function returns the physical value of the parameter
\begin{equation}\label{exactparam}
\rho = \fullint e_{\hat{p}}(t) x(t;\rho) dt.
\end{equation}

Nearly any time domain filter $f_{\hat{p}}(t)$ can be normalized to create a valid estimating function by requiring that the relation (\ref{exactparam}) be satisfied.  Thus, for linear parameters, the filter $f_{\hat{p}}(t)$ gives the estimating function
\begin{equation}
e_{\hat{p}}(t) \equiv \frac{f_{\hat{p}}(t)}{\int_{-\infty}^{\infty} f_{\hat{p}}(t) x(t;1) dt}
\label{estimatingfunc}
\end{equation}
where $x(t;1)$ is the unit-amplitude signal.  The only restriction on $f_{\hat{p}}(t)$ is that must not be orthogonal to the signal.

\subsection{\label{varwithspd}Calculating the Variance of Linear Parameter Estimates}

The variance in a parameter estimator is found using the estimating function and the autocovariance operator
\begin{equation}
\begin{split}
\text{var}(\hat{P}) = \int_{0}^{\tau} \int_{0}^{\tau} e_{\hat{p}}(t_1) \left< \delta X(t_1) \delta X(t_2) \right> e_{\hat{p}}(t_2) dt_1 dt_2,
\label{vartime}
\end{split}
\end{equation}
where we recall that capital letters represent ensembles.  This time-domain representation, however, is not necessarily the most convenient or intuitively appealing formulation of the variance.  For some noise processes, the Fourier basis is superior, yielding the expression
\begin{equation}
\text{var}(\hat{P}) = \frac{1}{2} \halfint F^2[e_{\hat{p}}(t);\nu] S[\delta X(t);\nu]d\nu,
\label{varfreq}
\end{equation}
where we denote functionals or linear operators with square brackets.  The spectral power density of the noise, $S[\delta X(t);\nu]$, is given by
\begin{equation}
S[\delta X(t);\nu] = 2 \fullint \left< \delta X(t) \delta X(t') \right> \cos (2 \pi \nu (t - t')) dt',
\label{spdwithautocovar}
\end{equation}
and $F^2[e_{\hat{p}}(t);\nu]$, which we call the Fourier energy density of the estimating function, is
\begin{equation}
F^2[e_{\hat{p}}(t);\nu] = \left( F_C[e_{\hat{p}}(t);\nu] \right)^2 + \left( F_S[e_{\hat{p}}(t);\nu]\right)^2,
\end{equation}
where
\begin{equation}
\begin{split}
& F_C[e_{\hat{p}}(t);\nu] = \sqrt{2} \fullint e_{\hat{p}}(t)\cos (2\pi \nu t)dt \\
& F_S[e_{\hat{p}}(t);\nu] = \sqrt{2} \fullint e_{\hat{p}}(t)\sin (2\pi \nu t)dt
\end{split}
\label{ftransforms}
\end{equation}
are the cosine and sine transforms of the estimating function.  We choose the normalization of (\ref{ftransforms}) to preserve Parseval's relation,
\begin{equation}
\halfint F^2[e_{\hat{p}}(t);\nu]d\nu = \fullint \left( e_{\hat{p}}(t) \right)^2dt.
\label{parseval}
\end{equation}
Throughout the remainder of this paper we will often drop the explicit dependence on $\nu$ of the Fourier energy density (FED) of the estimating function and the spectral power density (SPD) of the noise process and the dependence on $t$ of the data and of the estimating function.

\subsection{Power Density of Stationary Noise Processes}

The construction of the SPD in (\ref{spdwithautocovar}) requires that the time-dependent autocovariance operator be known \textit{a priori}.  In many instances a construction that uses the spectral information about a noise process is more transparent.  We present and discuss our preferred definition of the noise SPD in this section.

The inverse transform of equation (\ref{spdwithautocovar}) is equal to the autocovariance operator
\begin{equation}
\halfint S[\delta X] \cos (2 \pi \nu (t - t')) d\nu = \left< \delta X(t) \delta X(t') \right>.
\end{equation}
The special case where $t=t'$ gives the instantaneous variance of the noise ensemble
\begin{equation}
\halfint S[\delta X] d\nu = \left< \left( \delta X(t) \right)^2 \right> = \text{var}(\delta X(t))
\label{instvar}
\end{equation}
and shows that the SPD characterizes the contribution to the estimator variance from the noise contained in each infinitesimal frequency bin.  Our preferred definition of the noise SPD is therefore
\begin{equation}
\begin{split}
S[\delta X; \nu] \equiv \lim_{\Delta \nu \rightarrow 0} \Delta \nu \Bigl< & \left(F_C[\delta X ; \nu \pm \Delta \nu/2] \right)^2 \\
& + \left(F_S[\delta X ; \nu \pm \Delta \nu/2] \right)^2 \Bigr> \\
 = \lim_{\Delta \nu \rightarrow 0} \Delta \nu \Bigl< & F^2[\delta X ; \nu \pm \Delta \nu/2] \Bigr>
\label{spectralpower}
\end{split}
\end{equation}
where
\begin{equation}
\begin{split}
F_C[\delta x(t) ; \nu \pm \Delta \nu/2] & \equiv \sqrt{2} \int_{-\frac{1}{2\Delta \nu}}^{\frac{1}{2\Delta \nu}} \delta x(t) \cos(2 \pi \nu t) dt \\
F_S[\delta x(t) ; \nu \pm \Delta \nu/2] & \equiv \sqrt{2} \int_{-\frac{1}{2\Delta \nu}}^{\frac{1}{2\Delta \nu}} \delta x(t) \sin(2 \pi \nu t) dt
\label{finiteband}
\end{split}
\end{equation}
are the finite bandwidth Fourier components of a noise realization.

The relations (\ref{finiteband}) are an essential aspect of a Fourier definition of the SPD because the Fourier components of a noise realization diverge as $1/\sqrt{\Delta \nu} \sim \sqrt{\tau}$ in the limit as $\Delta \nu \rightarrow 0$ for a continuous noise spectrum.  Furthermore, if a realization of the noise were used in (\ref{spectralpower}) instead of the ensemble, $S[\delta x;\nu] = \lim_{\Delta \nu \rightarrow 0} \Delta \nu F^2[\delta x ; \nu \pm \Delta \nu/2]$, while remaining finite, would not converge.  Thus, the SPD is a property of the noise ensemble, not of any particular noise realization.

\subsection{Examples of Stationary Noise Processes}

We now calculate the variance in a parameter estimator due to the influence of three distinct stationary noise processes, monochromatic noise, white noise, and thermal noise.  We consider in detail the monochromatic case because it gives insight to parameter estimation using Fourier techniques and because it provides a straightforward way to verify the mutual consistency of the coefficients in the variance equation (\ref{varfreq}), in the two definitions of the SPD (\ref{spdwithautocovar}) and (\ref{spectralpower}), and in the normalization of the Fourier transforms (\ref{ftransforms}) and (\ref{finiteband}).

\subsubsection{Monochromatic Noise}

Consider the effect of an additive monochromatic noise component with amplitude $\epsilon$, frequency $\nu_0$, and random phase $\phi$ on a physical signal.  A realization of the data is
\begin{equation}
x(t) = x(t;\rho) + \epsilon \cos(2 \pi \nu_0 t - \phi).
\end{equation}
The parameter estimate is then
\begin{equation}
\begin{split}
\hat{p} = & \fullint e_{\hat{p}} x dt \\
= & \fullint e_{\hat{p}} \left( x(t;\rho) + \epsilon \cos(2 \pi \nu_0 t - \phi ) \right) dt\\
= & \ \rho + \frac{\epsilon}{\sqrt{2}} F_C[e_{\hat{p}};\nu_0] \cos \phi + \frac{\epsilon}{\sqrt{2}} F_S[e_{\hat{p}};\nu_0] \sin \phi,
\label{paramestimatesimple}
\end{split}
\end{equation}
an ensemble of which (each with random phase) constitutes the parameter estimator.  The variance of this estimator follows
\begin{equation}
\begin{split}
\text{var}&(\hat{P}_s) = \langle (\hat{P}_s - \rho)^2 \rangle \\
&= \frac{1}{2 \pi} \frac{\epsilon^2}{2} \int_0^{2 \pi} \left( F_C[e_{\hat{p}};\nu_0] \cos \phi + F_S[e_{\hat{p}};\nu_0] \sin \phi \right)^2 d\phi \\
&= \frac{\epsilon^2}{4} F^2[e_{\hat{p}};\nu_0],
\label{simplenoisevar}
\end{split}
\end{equation}
where the $s$ subscript denotes the single frequency noise model.

The variance in the ensemble of noise realizations is
\begin{equation}
\text{var}(\delta X_s) = \left< \left(\epsilon \cos(2 \pi \nu_0 t - \Phi) \right)^2 \right> = \frac{\epsilon^2}{2}.
\label{varmono}
\end{equation}
A substitution of (\ref{varmono}) into (\ref{instvar}) requires the monochromatic SPD to be
\begin{equation}
S[\delta X_s;\nu]=\frac{\epsilon^2}{2}\delta (\nu - \nu_0).
\label{monopower}
\end{equation}
This result, when substituted into equation (\ref{varfreq}), gives the same value for the parameter variance as the direct calculation, (\ref{simplenoisevar}), and shows that the coefficient of $1/2$ in (\ref{varfreq}) is consistent with our normalization convention for the Fourier transforms, (\ref{ftransforms}) and (\ref{finiteband}).

A rigorous derivation of the monochromatic SPD, using the definition (\ref{spectralpower}), gives the same result as (\ref{monopower})
\begin{equation}
\begin{split}
&S[\delta X_s;\nu]=\lim_{\Delta\nu\rightarrow 0}\Delta \nu \left< F^2[\delta X_s;\nu \pm \Delta\nu/2]\right> \\
&=\frac{\epsilon^2}{2} \\
&\times \lim_{\Delta\nu\rightarrow 0}\frac{\Delta\nu(\nu^2+\nu_0^2)\left( \sin^2 \left( \frac{\pi(\nu-\nu_0)}{\Delta\nu} \right) + \sin^2 \left( \frac{\pi(\nu+\nu_0)}{\Delta\nu} \right ) \right)}{\pi^2(\nu-\nu_0)^2(\nu+\nu_0)^2}\\
&=\frac{\epsilon^2}{2}\left( \delta(\nu-\nu_0)+\delta(\nu+\nu_0)\right)\\
&=\frac{\epsilon^2}{2}\delta(\nu-\nu_0)
\end{split}
\end{equation}
where the second term, $\delta(\nu+\nu_0)$, was dropped because the convention adopted in this paper does not allow negative frequencies.  This derivation shows the consistency of the two definitions of the SPD, (\ref{spdwithautocovar}) and (\ref{spectralpower}).

\subsubsection{White Noise}

Another example of a stationary noise process is white noise, with equal power at all frequencies,
\begin{equation}
S[\delta X_{wh}] = constant = \eta.
\end{equation}
This eliminates all two-point time correlations giving the autocovariance operator
\begin{equation}
\bigl< \delta X_{wh}(t_1) \delta X_{wh}(t_2) \bigr> = \frac{\eta}{2}\delta(t_1-t_2).
\label{whitenoisecovar}
\end{equation}
Although ideal white noise yields infinite power, for this paper we restrict our attention to calculations for which no non-physical results occur for pure white noise.  The variance in a parameter estimator may be written as
\begin{equation}
\begin{split}
\text{var}(\hat{P}_{wh}) & = \frac{1}{2} \int_{0}^{\infty} F^2[e_{\hat{p}}] S[\delta X_{wh}] d\nu \\
& = \frac{\eta}{2} \int_{0}^{\infty} F^2[e_{\hat{p}}] d\nu \\
& = \frac{\eta}{2} \int_{-\infty}^{\infty} \bigl( e_{\hat{p}} \bigr)^2 dt,
\end{split}
\end{equation}
where Parseval's relation was invoked in the final step.

\subsubsection{Thermal Noise}

The spectral power density of thermal noise is obtained beginning with the full equation of motion of the oscillator
\begin{equation}
\left( m \frac{d^2}{d t^2} + \xi \frac{d}{d t} + \kappa \right) X(t)=\mc{F}(t)
\end{equation}
where $\mc{F}(t)$ is the thermal driving force.  According to the fluctuation dissipation theorem, the SPD of the driving force associated with the thermal bath is a constant $4 k_B T \xi$~\cite{ca1,ca2}.  The response of the oscillator to this white driving force gives the SPD in displacement (cf. (\ref{specpow}))
\begin{equation}
S[\delta X_{th}] = \frac{4 k_B T \xi}{m^2 \left( \left( (2 \pi \nu)^2 - \omega_0^2 \right)^2 + (4 \pi \gamma \nu)^2 \right) }
\label{powerdensity}
\end{equation}
The autocovariance operator is (cf. (\ref{autocovar})).  
\begin{equation}
\begin{split}
& \left< \delta X_{th}(t) \delta X_{th}(t+\Delta t) \right>\\
&\quad =\frac{k_bTe^{-\gamma |\Delta t|}}{\kappa}\left(\cos(\omega|\Delta t|)+\left(\frac{\gamma}{\omega} \right)\sin(\omega|\Delta t|) \right)
\end{split}
\end{equation}
The integral of the displacement SPD for thermal noise, unlike the total power of displacement white noise is finite and is related to the expectation value for the potential energy of the oscillator
\begin{equation}
\text{var}(\delta X_{th}) = \int_{0}^{\infty} S[\delta X_{th}] d\nu = \frac{2 \langle P.E. \rangle }{\kappa} = \frac{k_B T}{\kappa} = \sigma^2.
\label{varxthermal}
\end{equation}
In section \ref{leucogenic} we find the optimal filter and calculate the parameter variance for a torsion balance in the presence of thermal noise.

\subsection{The Structure of Figure \ref{figure01}}

We now have the tools needed to describe the structure of Figure \ref{figure01}, the variance in the boxcar estimator of the equilibrium displacement of the oscillator as a function of the sample time.  The estimating function for the boxcar is
\begin{equation}
e_{\hat{c}}^{owh}(t) = \Theta(t;0,\tau)\frac{1}{\tau}
\end{equation}
where $\Theta(t;t_1,t_2) \equiv \theta(t-t_1) - \theta(t-t_2)$ is the boxcar function and $\theta(t)$ is the Heavyside (step) function.  The square of the Fourier transform of this estimating function gives the FED, $2\, \text{sinc}^2(\pi \nu \tau)$,
\begin{equation}
F^2[e_{\hat{c}}^{owh};\nu]=2\left(\frac{\sin(\pi\nu\tau)}{\pi\nu\tau}\right)^2.
\label{sincfunc}
\end{equation}
The SPD of thermal noise acting on the oscillator is a Lorentzian (\ref{powerdensity}), which for a high $Q$ oscillator peaks sharply near the resonance frequency.

\begin{figure}
\begin{center}
\includegraphics[width=\fsize]{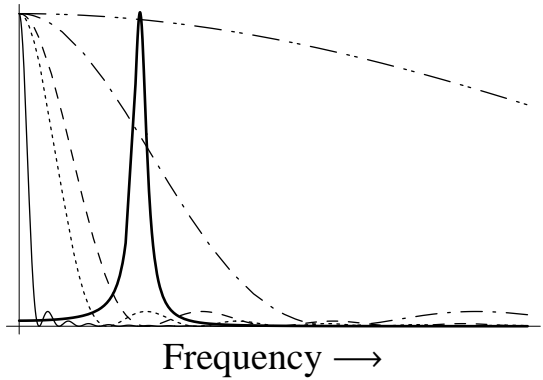}
\caption{Cartoon of a Lorentzian and several sinc$^2(\pi\nu\tau)$ FED curves, each corresponding to a different observation time, as a function of frequency.  The thick solid curve is a Lorentzian peak with $Q_0=7.5$ (chosen for cosmetic reasons).  The dash-two-dot FED curve corresponds to an observation time of 1/10 of a period, the dash-dot curve is for 1/2 a period, the dashed curve is for 1.25 periods, the dotted is for 1.8 periods, and the thin-line solid curve is for 8 periods.  We see that, while the 1.25 period observation nulls much of the noise from the Lorentzian peak, the 1.8 period observation allows a contribution from the peak.  The 8 period observation yields no significant contribution from the peak as the largest relative maxima of the FED have already passed the peak and instead allow contributions from the constant portion of the Lorentzian.  \label{lorandsinc}}
\end{center}
\end{figure}

When viewed as a function of frequency, the relative maxima and minima of the FED become more densely spaced as the observation time $\tau$ grows.  In contrast, the Lorentzian peak does not depend upon the observation time and remains fixed.  Since the variance (\ref{varfreq}) is proportional to the integral of the product of the SPD and the FED, the portions of the SPD near the minima of the FED contribute very little while the portions near the maxima, particularly the central maximum, contribute more significantly.  Figure \ref{lorandsinc} is a cartoon of a sample Lorentzian peak and several FEDs, each with different sample times, as a function of frequency.

We see that for very short sample times all of the Lorentzian noise contributes in nearly equal amounts because the sinc$^2(\pi \nu \tau)$ envelope is nearly constant.  As the sample time increases, the high frequency noise contributes less to the variance.  Eventually, the relative minima and maxima of the FED pass through the Lorentzian peak and the corresponding relative minima and maxima of Figure \ref{figure01} occur (primarily between 1 and 1/$Q_0$ periods).  Finally, for sufficiently long sample times, the central peak and several relative maxima of the FED are so close to the origin that the constant, low-frequency portion of the Lorentzian dominates the variance.

\section{\label{derivation}Derivation of the Optimal Filters}

Given that nearly any filter can be normalized to create an estimating function, we use the calculus of variation to find the optimal filter, $f_{\hat{p}}^{op}(t)$, that provides the minimum variance, unbiased estimator.  This procedure requires constraints on the duration of the data sample that are difficult to express in the Fourier basis.  As will be seen below, working in the time domain overcomes this challenge for white noise.  In section \ref{leucogenic} we address this problem for thermal noise.

\subsection{Optimal Filters for White Noise}

Consider the variance in a parameter estimate for the oscillator in the presence of white noise and with data from a continuous time series of length $\tau$,
\begin{equation}
\text{var}(\hat{P}_{wh}) = \frac{\eta}{2} \int_{0}^{\tau} e_{\hat{p}}^2 dt.
\end{equation}
We take the variation in this equation under the constraint (\ref{exactparam}), imposed by introducing a Lagrange multiplier $\lambda$, to obtain
\begin{equation}
\begin{split}
\delta \bigl(\text{var}&(\hat{P}_{wh})\bigr) \\
& = \delta \left( \frac{\eta}{2} \int_{0}^{\tau} e_{\hat{p}}^2 dt - \lambda \left( \int_{0}^{\tau} e_{\hat{p}} x(t;\rho) dt - \rho \right) \right) \\
& = \int_{0}^{\tau} \left( \eta e_{\hat{p}} - \lambda x(t;\rho) \right) \delta e_{\hat{p}} dt.
\end{split}
\end{equation}
Requiring that this variation be zero for all time shows that the optimal estimating function is proportional to the signal, $x(t;\rho)$.

A common expression of this filter, often called the ``matched'' filter, has the form of the unit-amplitude signal
\begin{equation}
f_{\hat{p}}^{owh}(t) \equiv \Theta (t;0,\tau)x(t;1).
\end{equation}
Normalizing the matched filter yields the optimal estimating function for white noise
\begin{equation}
e_{\hat{p}}^{owh} = \frac{\Theta(t;0,\tau) x(t;1)}{\int_{0}^{\tau} \left( x(t;1)\right) ^2 dt}.
\end{equation}
The boxcar function, $\Theta$, must be included in order to calculate the FED of the estimating function when working in the frequency domain.

\subsection{\label{leucogenic}Optimal Filters for Thermal Noise}

Since the autocovariance operator for thermal noise (\ref{autocovar}) is not diagonal, the optimal filter for thermal noise is more challenging to find, as the operator must be diagonalized.  To accomplish this, we first apply the equation-of-motion operator $\Omega$ to the data to obtain the thermal driving force
\begin{equation}
\Omega [x(t)] = \left( m \frac{d^2}{dt^2} + \xi \frac{d}{dt} + \kappa \right) x(t).
\label{eqnofmotion}
\end{equation}
Because the driving force $\mc{F}=\Omega[X_{th}(t)]$ is a white noise process with spectral power density $4k_BT\xi$, it has the same diagonal covariance operator as white noise (\ref{whitenoisecovar}) but with $\eta$ replaced by $4k_BT\xi$:
\begin{equation}
\begin{split}
\left< \Omega[X_{th}(t_1)] \Omega[X_{th}(t_2)] \right> & = \langle \Omega[\delta X_{th}(t_1)] \Omega[\delta X_{th}(t_2)] \rangle \\
& = 2k_B T \xi \delta(t_2 - t_1).
\end{split}
\end{equation}

Thus, when working in the acceleration basis (the basis of the thermal driving force), the matched filter provides the miminum variance estimator.  We define
\begin{equation}\label{accmatched}
z_{\hat{p}}(t) \equiv \Theta(t;0,\tau)\Omega[x(t;1)].
\end{equation}
to be the matched filter in the acceleration basis.  Note that the transformation to the acceleration basis removes information about the boundary conditions.  This loss of information is considered later.  We normalize and apply (\ref{accmatched}) to the stochastic driving force to find the parameter estimate
\begin{equation}\label{forcenormal}
\hat{p} = \frac{\fullint z_{\hat{p}} \Omega[x(t)] dt}{\fullint z_{\hat{p}} \Omega[x(t;1)] dt}.
\end{equation}

We now use $z_{\hat{p}}(t)$ to find a corresponding filter $f_{\hat{p}}(t)$ that can be applied directly to the displacement data by requiring
\begin{equation}
\fullint f_{\hat{p}}xdt = \fullint z_{\hat{p}}\Omega[x]dt
\end{equation}
for all realizations of $x$.  Integrating by parts yields the solution,
\begin{equation}
\begin{split}
&\fullint z_{\hat{p}} \Omega[x]dt = \fullint z_{\hat{p}} \left( m \ddot{x} + \xi \dot{x} + \kappa x \right) dt \\
&= \biggl[ z_{\hat{p}} \left( m \dot{x} + \xi x \right) \biggr]_{-\infty}^{\infty} - \fullint \left( \dot{z}_{\hat{p}} (m \dot{x} + \xi x) + \kappa z_{\hat{p}} x \right) dt \\
& = 0 - \biggl[ m \dot{z}_{\hat{p}} x \biggr]_{-\infty}^{\infty} + \fullint \left( m \ddot{z}_{\hat{p}} - \xi \dot{z}_{\hat{p}} + \kappa z_{\hat{p}} \right) x dt \\
& = 0 + \fullint \Omega^T[z_{\hat{p}}] x dt,
\end{split}
\end{equation}
where we introduce the transpose equation-of-motion operator
\begin{equation}
\Omega^T = m \frac{d^2}{dt^2} - \xi \frac{d}{dt} + \kappa.
\end{equation}
Thus, the optimal filter for thermal noise in the displacement basis obtained by a transformation from the acceleration basis is
\begin{equation}
f_{\hat{p}}^{oa}(t) = \Omega^T \left[ z_{\hat{p}}(t) \right] = \Omega^T \left[ \Theta(t;0,\tau) \Omega \left[ x(t;1) \right] \right]
\end{equation}
where the $oa$ superscript denotes that it is the \textit{optimal acceleration} filter.  Since $\Omega^T$ acts on the Heavyside functions, $f_{\hat{p}}^{oa}(t)$ can contain terms involving Dirac delta functions and their derivatives.  Normalizing this filter gives the optimal estimating function 
\begin{equation}
e_{\hat{p}}^{oa}(t) = \frac{\Omega^T \left[ \Theta(t;0,\tau) \Omega\left[x(t;1)\right] \right] }{\int_{-\epsilon}^{\tau+\epsilon} x(t;1) \Omega^T \left[ \Theta(t;0,\tau) \Omega\left[x(t;1)\right] \right] dt }
\end{equation}
where we include the infinitesimal $\epsilon$ to avoid ambiguity regarding how the denominator is evaluated.

\subsection{\label{opthcwg}Estimating the Equilibrium Displacement}

We now calculate the optimal filter and resulting parameter estimate for the equilibrium displacement of a thermally perturbed oscillator.  The results of this section are valid only for estimating a single parameter.  A subsequent paper will cover the more general case of several parameters.  We assume that the viscous drag coefficient $\xi$ and the torsional spring constant $\kappa$, or equivalently the damping coefficient $\gamma$ and the frequency $\omega_0$, are known.

The optimal filter for $c$ is
\begin{equation}\label{notquitefilt}
\begin{split}
f_{\hat{c}}^{oa}(t;c) =\ &\Omega^T \left[ \Theta(t;0,\tau) \Omega\left[x(t;1)\right] \right] \\
= m^2 \omega_0^4 & \biggl( (\theta(t) - \theta(t-\tau)) - 2 \frac{\gamma}{\omega_0^2}(\delta(t) - \delta(t-\tau)) \\
&+ \frac{1}{\omega_0^2}(\delta'(t) - \delta'(t-\tau)) \biggr)
\end{split}
\end{equation}
where $\delta '(t - t_0)$ is the time derivative of the delta function.  We normalize this filter to obtain the optimal estimating function for the equilibrium displacement of the oscillator,
\begin{equation}
e_{\hat{c}}^{oa}(t)=\frac{f_{\hat{c}}^{oa}(t;c)}{m^2 \omega_0^4 \tau},
\label{ofoestfunc}
\end{equation}
which yields the parameter estimate
\begin{equation}
\begin{split}
\hat{c}^{oa} & = \int_{0}^{\tau} e_{\hat{c}}^{oa} x(t) dt \\
& = \frac{x_f - x_i + Q_0 \omega_0 \tau x_m + Q_0\left(\frac{v_f - v_i}{\omega_0}\right)}{Q_0 \omega_0 \tau}.
\label{optaccel}
\end{split}
\end{equation}

The variance of the estimator corresponding to (\ref{optaccel}) is easiest to calculate in the acceleration basis.  A properly normalized filter must satisfy
\begin{equation}
\hat{c}^{oa} = \fullint e_{\hat{c}}^{oa} x dt = \fullint y_{\hat{c}}^{oa} \Omega \left[ x \right] dt
\label{estimateequiv}
\end{equation}
where
\begin{equation}
y_{\hat{c}}^{oa}(t) = \frac{\Theta(t;0,\tau)}{m\omega_0^2 \tau}
\end{equation}
is the result of normalizing the acceleration basis filter $z_{\hat{c}}$ as in (\ref{estimatingfunc}) replacing $x(t;1)$ with $\Omega[x(t;1)]$ .  The variance in the estimator $\hat{C}^{oa}$ is given by
\begin{equation}
\begin{split}
\text{var}(\hat{C}^{oa}) &= \frac{1}{2}\halfint F^2[y] S[\Omega[\delta X]] d\nu \\
&= 2 k_B T \xi \fullint y^2 dt \\
&= \frac{2\sigma^2}{Q_0 \omega_0 \tau}.
\end{split}
\end{equation}
\begin{figure}
\begin{center}
\includegraphics[width=\fsize]{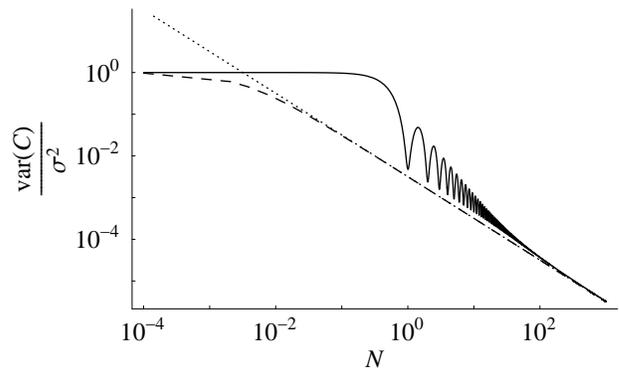}
\caption{The variance of the estimator $\hat{C}^{oa}$.  The solid curve is using the boxcar estimate (\ref{cboxcar}), the dotted line is using the optimal acceleration-only estimate (\ref{optaccel}), and the dashed line is the optimal estimate (\ref{copth}).}
\label{accelonlyplot}
\end{center}
\end{figure}

The dotted line in Figure \ref{accelonlyplot} shows the variance in the estimator $\hat{C}^{oa}$ for $Q_0=50$.  We see that this variance is indeed monotonic and smaller than the variance from the boxcar estimate for sample times larger than about 0.01 periods, however, for very short sample times the boxcar estimate has the smaller variance.  This failure results from the loss of information about the boundary conditions when transforming to the acceleration basis as mentioned in section \ref{leucogenic}.  These boundary conditions, when properly accounted for, rectify the failure of this approach for small sample times.

The \textit{initial} conditions are the natural boundary conditions because causality dictates that they depend only on the forces acting prior to the beginning of the sample.  Moreover, because the driving force on the oscillator is white, the force time series before the sample is uncorrelated with that during or after.  The initial displacement, initial velocity, and the acting forces completely determine the displacement of the oscillator.  We, therefore, write the optimal parameter estimate as linear combination of the initial conditions and a (not necessarily optimal) acceleration estimate
\begin{equation}\label{cwithweight}
\hat{c}^{oth} = w_1 x_i + w_2 \frac{v_i}{\omega_0} + w_3 \hat{c}^{a}.
\end{equation}
We wish to determine the choice of the constants $w_1$, $w_2$, and $w_3$ as well as the acceleration estimate that will produce the overall minimum variance, unbiased estimator.

Because the initial velocity contains no information about the equilibrium displacement, $w_2$ must be zero to minimize the variance in the estimator that corresponds to the parameter estimate (\ref{cwithweight}).  The variance is therefore
\begin{equation}\label{varcwithweight}
\text{var}(\hat{C})=w_1^2 \text{var}(X_i) + w_3^2  \text{var}(\hat{C}^{a}).
\end{equation}
The condition that the estimator be unbiased provides the constraint
\begin{equation}\label{const}
w_1+w_3=1
\end{equation}
and minimizing the total variance establishes that the weights $w_1$ and $w_3$ are proportional to the inverse variances.  The weights are therefore given by
\begin{equation}\label{weightinitial}
w_1 = \frac{\text{var}(\hat{C}^a)}{\text{var}(X_i) + \text{var}(\hat{C}^{a})}
\end{equation}
and
\begin{equation}\label{weightforceonly}
w_3 = \frac{\text{var}(X_i)}{\text{var}(X_i) + \text{var}(\hat{C}^{a})}
\end{equation}
and the variance simplifies to
\begin{equation}
\text{var}(\hat{C}) = \frac{\text{var}(X_i)\text{var}(\hat{C}^{a})}{\text{var}(X_i)+\text{var}(\hat{C}^{a})}.
\end{equation}
This last expression is minimized when the optimal acceleration estimator, described previously, is used for $\hat{C}^a$.

The optimal parameter estimate is then
\begin{equation}\label{optimal}
\begin{split}
\hat{c}^{oth} & = \frac{x_i \text{var}(\hat{C}^{oa})+\hat{c}^{oa}\text{var}(X_i)}{\text{var}(X_i)+\text{var}(\hat{C}^{oa})}\\
& = \frac{x_i+x_f+Q_0\omega_0\tau x_m + Q_0\left(\frac{v_f-v_i}{\omega_0}\right)}{2+Q_0\omega_0\tau}.
\end{split}
\end{equation}
Note that this parameter estimate has the same time-reversal symmetry as both the noise and the signal---a property that the acceleration-only estimate (\ref{optaccel}) does not share---and that the weight assigned to the initial displacement (\ref{weightinitial}) is that which restores the symmetry.  The variance in the optimal parameter estimator is
\begin{equation}\label{varoptimal}
\text{var}(\hat{C}^{oth}) = \frac{2 \sigma^2}{2 + Q_0 \omega_0 \tau}.
\end{equation}
The dashed curve in Figure \ref{accelonlyplot} shows the optimal variance as a function of the sample time duration.  For short time scales, the variance is constant and is dominated by the uncertainty in determining the initial displacement (\ref{cinst}).  For long time scales, the variance falls as $1/\tau$ and is dominated by the fluctuations induced by the thermal bath.

This behavior has implications for the utility of using an active feedback mechanism to damp the oscillator in an effort to reduce the total variance of the parameter estimator.  While damping the motion of the oscillator does indeed reduce the variance in the estimate of the initial displacement, it does not change the variance due to thermal excitiations because the thermal driving force depends solely upon the temperature of the environment.  Consequently, the value of using a feedback system depends upon the relative importance of the instantaneous measurement and the acceleration measurement of the equilibrium displacement for a particular experiment.  In many instances only the acceleration estimator is used and a feedback mechanism provides no benefit.

\section{\label{physical}Multiple Noise Processes}

The noise background of a physical system is generally a superposition of several noise processes.  Such a combination renders the task of finding the optimal estimator difficult if not impossible because, among other things, the basis in which the noise SPD is diagonal is unknown.  We investigate the effects that superposed white noise or residual transients caused by random, large amplitude disturbances to the oscillator have on the variance of several estimators: the boxcar, optimal thermal, and optimal acceleration estimators, as well as one that we will call the \EW\ (\ew) estimator.  The \ew\ estimator is related to the one used by the \EW\ experimental gravity group at the University of Washington~\cite{ade}.  Since the \EW\ group modulates their signal, their model involves several parameters.  Multi-parameter estimation and a detailed analysis of the estimator used by the \EW\ group will be covered in a subsequent paper.

\subsection{Transients}

Nonthermal disturbances to a high $Q$ oscillator may prevent the oscillator from ever reaching equilibrium with the thermal bath since the relaxation time of transients may be longer than the average time between the disturbances.  Because of this, the inclusion of the initial, instantaneous displacement estimate, $x_i$, in the optimal estimate (\ref{optimal}) can cause an increase in the variance of the estimator.  To overcome this, consider the optimal acceleration estimate in the acceleration basis (\ref{estimateequiv}) where the data time series is solely a transient
\begin{equation}
\begin{split}
c^{oa} 
&= \int_0^{\tau} y_{\hat{c}}^{oa} \Omega \left[ ae^{-\gamma t}\cos(\omega t)+be^{-\gamma t}\sin(\omega t) \right]dt = 0.
\end{split}
\end{equation}
Consequently, no transient can contribute to the parameter estimate or the estimator variance when using the acceleration estimate provided that the disturbance that causes the transient does not occur while the data are being aquired.

This shows that the acceleration estimator is superior under certain conditions.  To find these conditions we calculate the variance in the two estimators when an ensemble of random-phase disturbances with maximum displacement amplitude $\varepsilon$ cause a transient.  The optimal thermal estimator has a variance of (cf. (\ref{varcwithweight}) and (\ref{weightinitial}))
\begin{equation}
\text{var}(\hat{C}^{oth}) = \frac{2\sigma^2}{2+Q_0\omega_0\tau}+\frac{\varepsilon^2}{2}\left(\frac{2}{2+Q_0\omega_0\tau}\right)^2.
\end{equation}
Under the same conditions, the optimal acceleration estimator has a variance of
\begin{equation}
\text{var}(\hat{C}^{oa})=\frac{2\sigma^2}{Q_0\omega_0\tau}.
\end{equation}
The acceleration estimate is superior when
\begin{equation}
\frac{\varepsilon^2}{2}>\sigma^2 \left( 1+\frac{2}{Q_0\omega_0\tau} \right).
\end{equation}
With a high $Q$ oscillator, we see that transients \textit{as small as the thermal disturbances} can render the optimal thermal estimate inferior to the acceleration estimate.  This is true even when the sample time is small ($\tau \sim 1/\omega_0 Q_0$).  For this reason, the acceleration estimator is often used in lieu of one that accounts for the initial displacement.

\subsection{White Noise and Thermal Noise Combined}

\subsubsection{\label{comparison}E\"ot-Wash Approach}

When additive white noise is present, the use of $x_i$, $x_f$, $v_i$, and $v_f$ in the optimal estimate yields infinite variance in the optimal thermal estimator.  However, if the white noise does not dominate, one need not resort to the boxcar estimate.  The \ew\ estimator is quite robust for systems that are dominated by either white noise or thermal noise.  Its variance is within approximately $1/N$ of the optimum in either case, where $N$ is the number of oscillation periods in the data sample.

\begin{figure}
\begin{center}
\includegraphics[width=\fsize]{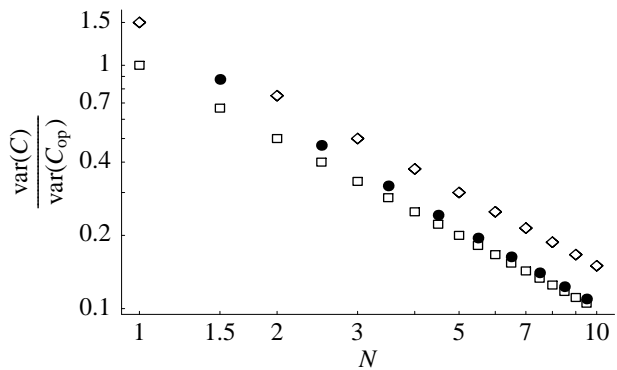}
\caption{A plot of the normalized variance of the boxcar (diamond), the \ew\ (filled circle), and the optimal (square) estimators in the presence of thermal noise as a function of sample time (in terms of the number of periods).  The results are calculated to first order in $1/Q_0$.}
\label{eotwashvariance}
\end{center}
\end{figure}

\begin{figure}
\begin{center}
\includegraphics[width=\fsize]{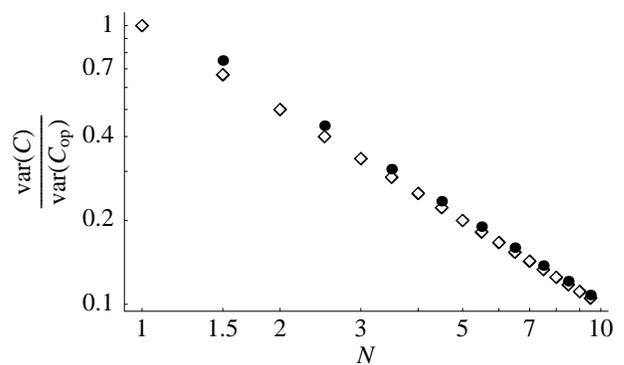}
\caption{A plot of the normalized variance of the \ew\ (filled circle) and boxcar (diamond) estimators in the presence of white noise as a function of the sample time in terms of the number of periods.}
\label{eotwashvsboxwhite}
\end{center}
\end{figure}

The \ew\ approach averages the data with itself delayed by half of a period.  A boxcar average is then taken for an integer number of oscillation periods.  The variance in the \ew\ estimator in the presence of purely thermal noise is less than that of the boxcar estimator.  Figure \ref{eotwashvariance} shows a comparison of the variance of the \ew\ estimator with that of the optimal thermal estimate and the boxcar estimate as a function of the sample time.
We see that the variance in the \ew\ estimator is situated between the boxcar and optimal estimates and it approaches the optimal as roughly $1/N$.  The robustness of the \ew\ estimator is manifest when we examine the variance of the same estimators in the presence of white noise.  Because the optimal thermal estimator has infinite variance for this case, we show in figure \ref{eotwashvsboxwhite} the variance in the \ew\ estimator compared with the variance of the boxcar in the presence of white noise as a function of the sample time.

Not only is the \ew\ estimator robust under these changes in the noise background, its variance is more immune by a factor of $1/Q_0^2$ to the transient signal than is the boxcar.  To illustrate this property, consider a boxcar estimate.  To leading order in $1/Q_0$ and for an integer number of periods
\begin{equation}
\begin{split}
\hat{c}^{owh}&=\frac{1}{n\mc{P}}\int_0^{n\mc{P}}x(t)dt\\
&=\frac{1}{n\mc{P}}\int_0^{n\mc{P}}(ae^{-\gamma t}\cos(\omega t)+be^{-\gamma t}\sin(\omega t) +c)dt\\
&=c+b\frac{\gamma}{\omega}+O\left(\frac{1}{Q_0}\right)^2
\end{split}
\end{equation}
where $\mc{P}$ is the period of the damped oscillator.  The variance in the boxcar estimator, expressed to the same order, is
\begin{equation}
\text{var}(\hat{C}^{owh})= \frac{3\sigma^2}{Q_0\omega \tau} + \frac{\varepsilon^2}{2}\left(\frac{1}{Q_0}\right)^2.
\end{equation}
In order for the fractional increase in variance to be small, the amplitude of the transient disturbance, $\varepsilon$, must satisfy
\begin{equation}\label{smallincbox}
\frac{\varepsilon^2}{2} \ll \frac{6Q_0\sigma^2}{n\mc{P}\pi}.
\end{equation}
By comparison, with an extra half-period of data, the \ew\ estimate is
\begin{equation}
\begin{split}
\hat{c}^{ew}&= \frac{1}{2n\mc{P}}\int_{0}^{n\mc{P}}x(t)dt + \frac{1}{2n\mc{P}}\int_{\mc{P}/2}^{(n+1/2)\mc{P}}x(t)dt\\
&=c+a\frac{\pi}{2}\left( \frac{\gamma}{\omega} \right)^2 + O\left( \frac{1}{Q_0} \right)^3
\end{split}
\end{equation}
and the variance of the estimator is
\begin{equation}
\text{var}(\hat{C}^{ew})=\frac{3\sigma^2}{Q_0\omega_0 \tau}+\frac{\varepsilon^2}{2}\frac{\pi^2}{4}\left(\frac{1}{Q_0}\right)^4.
\end{equation}
In this case, a small increase in variance need only satisfy
\begin{equation}
\frac{\varepsilon^2}{2} \ll \frac{96Q_0^3\sigma^2}{n\mc{P}\pi^3},
\end{equation}
a significant relaxation of the constraint for the boxcar, (\ref{smallincbox}).

\subsubsection{Numerical Results}

Because it is generally difficult to transform to a representation in which an arbitrary mixture of noise has a white power spectrum, numerical methods are often the only option available to reduce the uncertainty in a measurement due to the estimation technique.  To employ numerical methods the data is discretized.  The optimal estimator is then found using generalized least-squares analysis~\cite{ham}.  As an example, we calculate the optimal estimator using one and one-half periods of data sampled at 300 points.  For a single linear parameter, the optimal parameter estimate is found using a discrete filter given by
\begin{equation}\label{discfilt}
\B{e}_{\hat{p}}^{op} = \left( \B{q}^T \ \mat{m}_X^{-1} \ \B{q} \right)^{-1} \ \B{q}^T \ \mat{m}_X^{-1}
\end{equation}
where $\mat{m}_X$ is the noise covariance matrix and $\B{q}$ is sometimes called the design vector.  The design vector is given by the partial derivative of the parameterized data with respect to the parameter at each time step
\begin{equation}
\B{q}=\frac{\partial}{\partial c} \B{x}.
\end{equation}
For the equilibrium displacement of the oscillator, each component of the design vector is unity.  The data is multiplied by the filter (\ref{discfilt}) to give the parameter estimate.

\begin{figure}
\begin{center}
\includegraphics[width=\fsize]{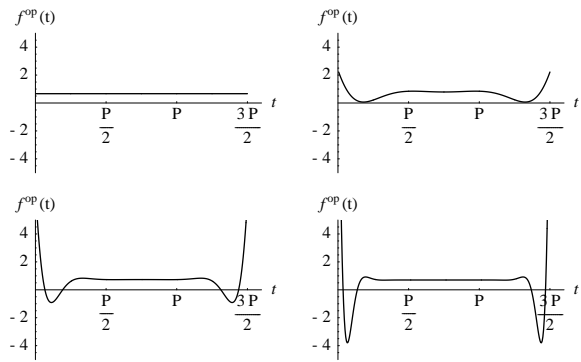}
\caption{Paneled plots of the optimal filters derived from 300 discrete datum for a mixture of white and thermal noise.  The top right corner is the optimal filter for pure white noise, top left has 10\% white, lower left has 1\% white, and lower right has 0.1\% white noise.}
\label{optdiscrete}
\end{center}
\end{figure}

To investigate the changes in the optimal filter as the noise background changes from pure white noise to pure thermal noise, we normalized the noise covariance matrices for white and thermal noise so that, with one and one half periods of data, the \ew\ estimator has unit variance.  We then combine some fraction of each of the covariance matrices so that the sum of the admixture coefficients is unity.  Figure \ref{optdiscrete} shows an interpolation of the optimal estimating vector for different mixtures of noise.  We see that the optimal filter starts as a boxcar for pure white noise and approaches the combination of a boxcar with Dirac delta function derivatives (\ref{optimal}) as the fraction of white noise is decreased.

We evaluated the variance of the optimal estimator and compared it with the unity variance of the \ew\ estimator for several noise mixtures.  For the case of pure white noise, the variance in the optimal estimator is 89\% of the variance in the \ew\ estimator.  The optimal estimator variance is 90\% of the \ew\ estimator variance for 10\% white noise, 84\% for 1\% white noise, and 80\% for 0.1\% white noise.  For pure thermal noise (not shown), the variance in the optimal estimator is 70\% of that in the \ew\ estimator.  This analysis is valid for a mixture of only white noise and thermal noise; transient signals were not included.  Filters such as those shown in figure \ref{optdiscrete} are not generally immune to transient signals.  This fact again illustrates the robustness of the \ew\ estimator because, in the variance, transient signals are only manifest at fourth order in $1/Q_0$.

\section{Discussion}

Equation (\ref{optimal}) defines the minimum variance, unbiased estimator for the equilibrium displacement of a damped harmonic oscillator when statistical fluctuations in thermal equilibrium are the dominant source of noise.  In deriving this estimate we chose to transform the observable to the acceleration basis in which the thermal noise spectral power density has a diagonal form (equal noise power at all frequencies).  Once in this ``white noise'' basis, the minimum variance estimator is determined by application of the matched filter.  A subsequent transformation of this estimator back into the displacement representation gives our result.

This \textit{closed-form} solution is of great advantage to the experimentalist.  Such a solution for any noise process serves to guide the design of an experimental apparatus and the methods used to gather and reduce the experimental data.  The corresponding solution for white noise, the boxcar estimator, has been used extensively as an optimal estimator under proper criteria or as a point of departure for the construction of an appropriate estimator.

One example is the \ew\ estimator which is robust and near optimum under the restriction that the data sample is a half-integer number of periods in duration.  In a laboratory such a requirement can often be met, but there are circumstances where this criteron is either inefficient, not feasable, or 
may be entirely beyond the control of the observer as is the case in relevant astrophysical scenarios.  In such situations the \ew\ estimator would fail to be near optimum and possibly fail to be defined (e.g. if only a single half period of data is given).  Since the \ew\ estimator does not generally apply, one might be tempted to resort to the boxcar estimator.  According to figure \ref{figure01} the penalty for such a choice can be an increase in variance by a factor of order $Q$.  Such an increase can occur when the assumptions implicit in formulating optimal estimates, like the boxcar and optimal thermal estimates, are not satisfied.

When both white and thermal noise processes are present, neither solution is appropriate.  Moreover, the method used in section \ref{derivation} to minimize the variance (transforming to a diagonal representation) may not be possible.  Under certain circumstances one may find an estimator that is relatively immune to combinations of noise, such as the \ew\ estimator.  More generally, the only practical option is to discretize the data and use least-squares methods to find the optimal estimator numerically.  In such situations, the interpretation of the numerical results may not be obvious and the closed form solution can provide appropriate guidance (c.f. figure \ref{optdiscrete}).

While we have addressed some aspects of \textit{random} noise beyond thermal noise, there are several \textit{systematic} effects that we have neglected.  These effects can be roughly divided into two groups: effects that can be modeled and incorporated into the analysis of the data and those that cannot.  The latter group, which includes such things as temperature fluctuations, fiber anelasticity or nonlinearity, and sudden relaxations of the fiber (fiber quakes), will not be discussed in our articles.  The former group, which includes linear fiber drift, damped oscillations, signal modulation, etc. we will discuss.  However, incorporating these effects into the analysis requires an extension of the techniques developed in this paper.  In future publications we will address simultaneous fitting for several \textit{linear} parameters (for example, to fit for a modulated signal or linear fiber drift) and \textit{nonlinear} parameters (such as the oscillation frequency and damping coefficient of the oscillator).

These subsequent papers will also discuss some of the implications that the analytic results have on experimental design.  We have already mentioned at the end of section \ref{derivation} that the use of active feedback to damp the motion of the oscillator when estimating the equilibrium displacement is beneficial only if one uses the instantaneous estimate (\ref{cinst}) when determining the equilibrium displacement of the oscillator---compare (\ref{optaccel}) and (\ref{optimal}).  Another striking fact is revealed when fitting for the oscillation frequency of the oscillator.  We will show that, for thermal noise, the optimal estimate of the oscillation frequency requires no more than four measurements of the displacment of the oscillator each period.  That is, there is no direct benefit from having five or more displacment measurements for thermal-noise-limited experiments where the oscillation frequency is the signal.  These two examples demonstrate how an analytic expression for optimal parameter estimators can have significant implications for the design of an experimental apparatus---insights that do not emerge from numerical solutions.

\begin{acknowledgments}
We would like to thank Dr. Brian Walton, Dr. Don Percival, and Dr. John Deeter for the useful discussions we had concerning this work and the NSF (Grant PHY-0244762) for partial support of this work.
\end{acknowledgments}


\end{document}